\documentclass[12pt,a4paper]{article}
\usepackage[dvips]{graphicx}
\renewcommand\theequation{\arabic{section}.\arabic{equation}}
\begin{document}

\title{Spinodal curve of a three-component molecular system}

\author{Florin D. Buzatu\thanks{On leave from Department of
Theoretical Physics, National Institute for Physics and Nuclear
Engineering, Bucharest-M\u{a}gurele, 76900, Romania.}
\thanks{Corresponding author.}, Daniela Buzatu\thanks{On leave from
Physics Department, Politehnica University, Bucharest, 77206,
\mbox{Romania}.}, \\
and \\ John G. Albright
\\ \textit{Department of Chemistry, Texas Christian University,} \\
\textit{Fort Worth, TX 76129, USA} }

\date{- March 14, 2001 -}
\maketitle

\begin{abstract}
We consider a lattice model for three-component systems in which
the lattice bonds are covered by molecules of type $AA$, $BB$, and
$AB$, and the only interactions are between the molecular ends of
a common lattice site. The model is equivalent to the standard
Ising model, and the coexistence curves for different lattices
and/or some specific form of the interactions have been previously
investigated. We derive the spinodal curve of the three-component
model on the honeycomb lattice based on the mean-field and
Bethe-lattice results of the equivalent Ising model. The spinodal
and the coexistence curves of the ternary solution are drawn at
different values of the reduced temperature, the only parameter of
the model. The particular case of a two-component system is also
illustrated.
\end{abstract}

\newpage
\section{Introduction}

\hspace{\parindent}Two almost completely immiscible liquids may
increase their mutual solubility in the presence of a solvent. By
increasing the solvent concentration, the two liquid layers become
more alike and, at some (temperature dependent) composition, a
single homogeneous phase is formed. In order to describe this
phenomenon, Wheeler and Widom \cite{WhW68} considered three types
of diatomic molecules, $AA$, $BB$, and $AB$, covering the bonds of
a regular lattice (one molecule per bond) and subject to the
condition that only $A$ atoms or only $B$ atoms may meet at a
given lattice site (infinite repulsion between $A$ and $B$ atoms
of a common site, no interaction otherwise). The model reduces to
the standard (spin-1/2, nearest-neighbor interaction) Ising model
on the same lattice and its ferromagnetic transition corresponds
to a phase-separation transition for the ternary solution.
Although the model catches the main aspects of the
phase-separation transition, it is not temperature dependent. The
model can be extended to include temperature as a significant
variable by allowing finite interactions between any two atoms of
a common site. This \emph{extended Wheeler-Widom model} is still
equivalent to the standard Ising model but on a decorated lattice
in which each site of the original lattice is replaced by a
cluster of sites. The exact coexistence (binodal) curves of this
extended Wheeler-Widom model on the honeycomb and Bethe lattices
were determined in \cite{HuS86} and \cite{HSB89}, respectively.
Further extensions of the model have been considered in
\cite{HuS89,HuS90,SHW91}.

A solution suddenly quenched from an initial one-phase state in
the two-phase region reaches its equilibrium through either a
nucleation or a spinodal decomposition process \cite{GSS83}. The
states below the coexistence curve can be then classified as
metastable (close to the binodal) and unstable (deep in the
two-phase region), respectively. The \emph{spinodal curve} is
supposed to separate these two types of behavior. According to the
classical (mean-field) theory of first-order phase transitions,
the spinodal is defined as the locus for which an appropriate
susceptibility diverges; for a multi-component solution, it
follows that the determinant of the diffusion coefficients must be
zero and this provides an experimental determination of the
spinodal curve. Although a mean-field-like approach oversimplifies
the problem, it is still a convenient way to distinguish between
the metastable and unstable states.

\begin{figure}[p]
\begin{center}
\includegraphics[]{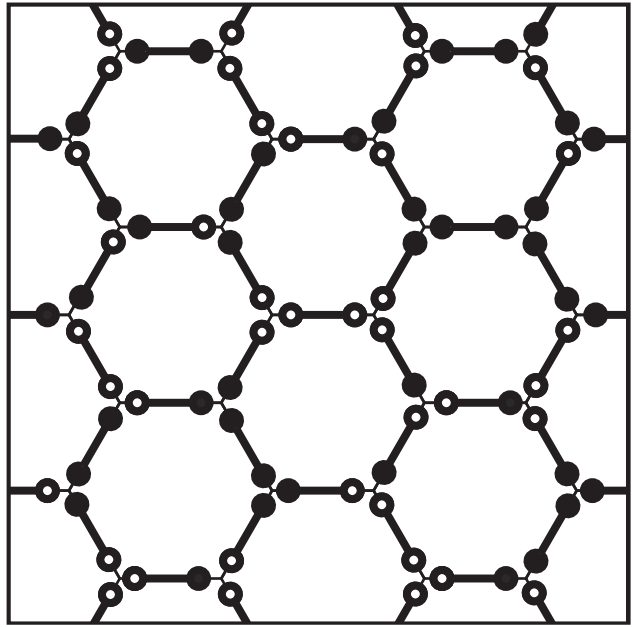}
\caption{Configuration of molecules within the extended
Wheeler-Widom model on the honeycomb lattice.} \label{fig:molec}
\end{center}
\end{figure}

\begin{figure}[p]
\begin{center}
\includegraphics[]{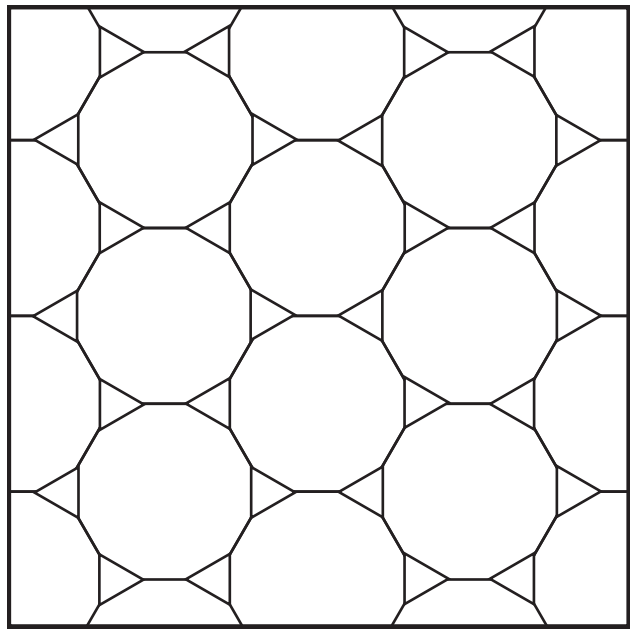}
\caption{A portion of the 3-12 lattice, each site being covered by
one $C_3$ graph (triangle) and one $C_2$ graph (segment connecting
two triangles).} \label{fig:3-12}
\end{center}
\end{figure}

The aim of the present paper is to determine the spinodal curve
for the extended Wheeler-Widom model. The molecules are considered
to occupy the bonds of a honeycomb lattice (see Fig.
\ref{fig:molec}) and the lattice of the corresponding Ising model
is called 3-12 (see Fig. \ref{fig:3-12}). As shown in Section 2,
by applying at first a star-triangle transformation and then a
double-decoration, the Ising model on the 3-12 lattice becomes the
Ising model on the \mbox{honeycomb} lattice, the original lattice
of the molecular model. In Section 3 we shortly review two
approximate treatments of the ferromagnetic Ising model:
\emph{mean-field} and \emph{Bethe-lattice}. The resulting
coexistence and spinodal curves for the three-component model, the
effect of temperature, and the limit of a two-component system are
all presented in Section 4. In the last section we discuss some
possible developments of the present work.

\section{Extended Wheeler-Widom model on the honeycomb lattice}
\setcounter{equation}{0}

\hspace{\parindent} The grand-canonical partition function of the
Wheeler-Widom model with finite interactions on the honeycomb
lattice ($N$ sites) is proportional to the partition function of
the standard Ising model on the 3-12 lattice \cite{HuS86}
\begin{equation}\label{eq:z312}
Z^{}_{\mbox{\scriptsize{3-12}}}=\sum_{\{\sigma_i\}}\exp\left(
R\sum_{(i,j)\subset C_3}\sigma_i\sigma_j + L\sum_{(i,j)\subset
C_2}\sigma_i\sigma_j + h\sum_{i}\sigma_i\right)
\end{equation}
with $\sigma_{i},\sigma_{j}=\pm 1$ ($i,j=1,\dots ,N$) depending on
the spin orientation. The first sum in (\ref{eq:z312}) is over all
possible configurations, the second one runs over all the
triangles ($C_{3}$ graphs) of the 3-12 lattice, the third one is
over all the bonds ($C_{2}$ graphs) connecting the triangles, and
the last one is over the whole lattice (see Fig. \ref{fig:3-12}).
The parameters $R$, $L$, and $h$ from (\ref{eq:z312}) are related
to the parameters of the original model as follows
\begin{equation}
 R=\frac{1}{4k_{B}T}\label{eq:r}
 (2\varepsilon_{AB}-\varepsilon_{AA}-\varepsilon_{BB}),
\end{equation}
\begin{equation}\label{eq:l}
 L=\frac{1}{4k_{B}T} (\mu_{AA}+\mu_{BB}-2\mu_{AB}),
\end{equation}
and
\begin{equation}\label{eq:h}
 h=\frac{1}{4k_{B}T}(\mu_{AA}-\mu_{BB}-2\varepsilon_{AA}+
 2\varepsilon_{BB}).
\end{equation}
Here $k_{B}$ is Boltzmann's constant and $T$ is the temperature.
We use $\varepsilon_{AA}$, $\varepsilon_{AB}$, and
$\varepsilon_{BB}$ to denote the interaction energies between the
molecular ends of a common lattice site. The chemical potentials
$\mu_{AA}$, $\mu_{BB}$, and $\mu_{AB}$ all tend to infinity (no
vacant sites are allowed), but differences of any two of them are
finite thermodynamic variables.

\begin{figure}[p]
\begin{center}
\includegraphics[]{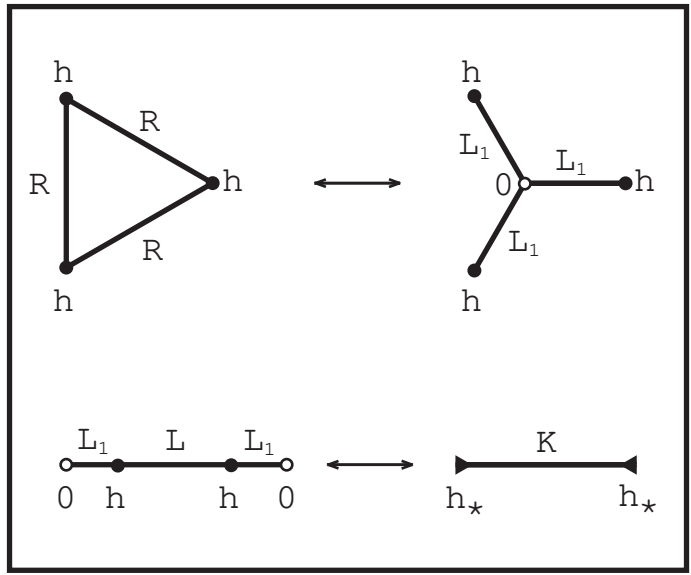}
\vspace{1cm} \caption{Star-triangle transformation (up) and
double-decoration (down) by which the Ising model on the 3-12
lattice can be transformed into the Ising model on the honeycomb
lattice. At constant $L_1$, the double-decoration can be also
inverted (i.e., from right to left).} \label{fig:trfs}
\end{center}
\end{figure}

The partition function of the standard Ising model on the 3-12
lattice can be connected to the partition function of the standard
Ising model on the honeycomb lattice
\begin{equation}\label{eq:zhon}
Z^{}_{\mathrm{hon}}=\sum_{\{\sigma_i\}}\exp\left(
K\sum_{<i,j>}\sigma_i\sigma_j + 3h_{\star}\sum_{i}\sigma_i\right)
\end{equation}
via two transformations: star-triangle followed by
double-decoration (see \mbox{Fig. \ref{fig:trfs}}). The result is
\cite{Huc86}
\begin{equation}\label{eq:z312hon}
Z^{}_{\mbox{\scriptsize{3-12}}}(R,L,h)=A^{-N}(L_{1})B^{3N/2}(L_{1},L,h)Z^{}_{\mathrm{hon}}(K,3h_{\star}),
\end{equation}
where
\begin{equation}\label{eq:a}
  A^{4}(L_{1})=16\cosh(3L_{1})\cosh^{3}(L_{1})
\end{equation}
and
\begin{equation}\label{eq:l1}
  \cosh(2L_{1})=\left(e^{4R}+1\right)/2 \ \ , \ \ R>0,
\end{equation}
are obtained from the star-triangle transformation, and
\begin{eqnarray}
 B^{4} &=& 16e^{-4L}\left[ 1+e^{2L}\cosh(2L_{1}+2h) \right] \left[
 1+e^{2L}\cosh(2L_{1}-2h) \right] \nonumber \\
    & & \times \left[ \cosh(2L_{1})+e^{2L}\cosh(2h) \right]^{2}, \label{eq:b}
\end{eqnarray}
\begin{equation}\label{eq:k}
 e^{4K}=\frac{\left[ 1+e^{2L}\cosh(2L_{1}+2h)
 \right] \left[ 1+e^{2L}\cosh(2L_{1}-2h) \right]}
 {\left[ \cosh(2L_{1})+e^{2L}\cosh(2h) \right]^{2}},
\end{equation}
and
\begin{equation}\label{eq:hs}
 e^{4h_{\star}}=\frac{1+e^{2L}\cosh(2L_{1}+2h)}
 {1+e^{2L}\cosh(2L_{1}-2h)}
\end{equation}
follow from the double-decoration transformation. The parameter
$L_1$ from equations (\ref{eq:z312hon}) to (\ref{eq:hs}) is an
interaction constant of the Ising model on the intermediate
lattice (see Fig. \ref{fig:trfs}); its sign is undetermined and,
without loss of generality, we may choose $L_1>0$.

Let us denote by $X_{AA}$, $X_{BB}$, and $X_{AB}$ the mole
fractions respectively corresponding to the three types of the
molecules. The mole fractions obey the conservation equation
\begin{equation}\label{eq:consv}
 X_{AA}+X_{BB}+X_{AB}=1
\end{equation}
and they can be related to the quantities of the equivalent Ising
model by \cite{HuS86}
\begin{equation}\label{eq:molfrsig}
 X_{AA}+X_{BB}-X_{AB}=\sigma^{}_{\mbox{\scriptsize{3-12}}}
\end{equation}
and
\begin{equation}\label{eq:molfrmag}
 X_{AA}-X_{BB}=m^{}_{\mbox{\scriptsize{3-12}}},
\end{equation}
where
$\sigma^{}_{\mbox{\scriptsize{3-12}}}=<\sigma_{i}\sigma_{j}>_{C_{2}}$
describes the spin-spin correlation on the $C_{2}$ graphs and
$m^{}_{\mbox{\scriptsize{3-12}}}=<\sigma_{i}>$ is the
magnetization of the 3-12 lattice. Using (\ref{eq:z312hon}),
$\sigma^{}_{\mbox{\scriptsize{3-12}}}$ and
$m^{}_{\mbox{\scriptsize{3-12}}}$ can be expressed in terms of the
similar quantities $\sigma$ and $m$ on the honeycomb lattice as
\begin{equation}\label{eq:sig312}
 \sigma^{}_{\mbox{\scriptsize{3-12}}}=\left(\frac{\partial \ln
 B}{\partial L}\right)_{L_{1},h} +\left(\frac{\partial K}{\partial
 L}\right)_{L_{1},h}\sigma+
 2\left(\frac{\partial h_{\star}}{\partial L}\right)_{L_{1},h}m
\end{equation}
and
\begin{equation}\label{eq:mag312}
 m^{}_{\mbox{\scriptsize{3-12}}}=\frac{1}{2}\left(\frac{\partial \ln
 B}{\partial h}\right)_{L_{1},L}+ \frac{1}{2}\left(\frac{\partial
 K}{\partial h}\right)_{L_{1},L}\sigma+ \left(\frac{\partial
 h_{\star}}{\partial h}\right)_{L_{1},L}m.
\end{equation}

The Ising model has two parameters, $K$ and $h^{}_{\star}$, while
the Wheeler-Widom model has three independent combinations of the
original parameters: $R$, $L$, and $h$. One can choose $1/R$ as
the \emph{reduced temperature} of the molecular model
\cite{HuS86}; we will use $t$ to denote it and keep it constant
throughout the calculation. (When $\varepsilon_{AB}\rightarrow
\infty$, then $t\rightarrow 0$ and the model reduces to the
original model introduced by Wheeler and Widom \cite{WhW68}; see
also Appendix.) Because we describe the molecular model by means
of the Ising model, all the derivatives from the equations
(\ref{eq:sig312}) and (\ref{eq:mag312}) must be expressed in terms
of $R$, $K$, and $h^{}_{\star}$ rather than $L_1$, $L$, and $h$.
The interaction constant $L_{1}$ as a function of $R$ is given by
(\ref{eq:l1}) and from (\ref{eq:k}) and (\ref{eq:hs}) we can find
$L$ and $h$ as functions of $K$ and $h^{}_{\star}$ at constant
$L_{1}$ (or, equivalently, at constant $t$). Namely,
\begin{equation}\label{eq:chi}
 \tanh(2h)=\frac{\sinh(2L_{1}) \sinh(2h^{}_{\star})}
 {\cosh(2L_{1}) \cosh(2h^{}_{\star})-e^{-2K}}\equiv \chi
\end{equation}
and
\begin{equation}\label{eq:lambda}
 \frac{e^{2L}}{\sqrt{1-\chi^{2}}}= \frac{\cosh(2L_{1})
 \cosh(2h^{}_{\star})-e^{-2K}}
 {\cosh(2L_{1}) e^{-2K}- \cosh(2h^{}_{\star})} \equiv \lambda.
\end{equation}
The expressions (\ref{eq:chi}) and (\ref{eq:lambda}) are well
defined only if $|\chi|<1$ and \mbox{$\lambda >0$}, which amounts
to
\begin{equation}\label{eq:hsrange}
 |h^{}_{\star}|< L_1
\end{equation}
and
\begin{equation}\label{eq:krange}
 -\frac{1}{2}\ln[ \cosh(2L_1-2|h^{}_{\star}|)] <
 K < \frac{1}{2}\ln\left[
 \frac{\cosh(2L_1)}{\cosh(2h^{}_{\star})}\right]\equiv K_{\max},
\end{equation}
the latter inequalities determining the region in the
$(K,h^{}_{\star})$-space which can be mapped into $(L,h;R)$-space.
This region becomes the whole space when $R\rightarrow \infty$.
The derivatives from (\ref{eq:sig312}) and (\ref{eq:mag312}) as
functions of $K$, $h_{\star}$, and $R$ are given in Appendix.
Provided $\sigma$ and $m$ (both functions of $K$ and $h_{\star}$)
are known, we have now the complete ``dictionary" between our
molecular system and the Ising model.

The ferromagnetic transition of the Ising model ($K>K_c>0$,
$h_{\star}=0$) corresponds to a phase-separation transition for
the Wheeler-Widom model ($L>0$, $h=0$). If $\sigma^{}_{0}(K)\equiv
\sigma(K,0)$ and the spontaneous magnetization $m^{}_{0}(K)\equiv
m(K,0)$ from (\ref{eq:sig312}) and (\ref{eq:mag312}) are
specified, then one can determine the isothermal coexistence
curves for our molecular system by eliminating $K$ between the
equations (\ref{eq:molfrsig}) and (\ref{eq:molfrmag}). The
coexistence curves based on the exact solution of the Ising model
on the honeycomb lattice and on the three-coordinated Bethe
lattice were determined in the \cite{HuS86} and \cite{HSB89},
respectively. In finding the spinodal curve, we need
$\sigma^{}_{s}(K)\equiv \sigma[K,h^{(s)}_{\star}(K)]$ and
$m^{}_{s}(K)\equiv m[K,h^{(s)}_{\star}(K)]$ calculated along a
certain path $h^{(s)}_{\star}(K)$ of the Ising model. We consider
two approximate methods below: mean-field and Bethe-lattice.

\section{Ising ferromagnet in the mean-field approximation and on the Bethe lattice}
\setcounter{equation}{0}

\hspace{\parindent}In the {\em mean-field approximation}, the
magnetization of the Ising mo\-del with the partition function
(\ref{eq:zhon}) is the solution of the equation \cite{Pat78}
\begin{equation}\label{eq:magmf}
  m=\tanh (qKm+3h^{}_{\star}),
\end{equation}
where $q$ is the coordination number of the lattice. This equation
is invariant under the transformation $(m,h^{}_{\star})\rightarrow
(-m,-h^{}_{\star})$, i.e., $h^{}_{\star}$ is an odd function of
$m$. The spin-spin correlation function $\sigma$ can be
determined, for example, by using its relation to the internal
energy per spin $u$,
\begin{equation}\label{eq:sigmf}
  \sigma =\frac{2}{qK}\left( \frac{u}{k_{B}T} +3h^{}_{\star}
  m\right),
\end{equation}
and substituting the expression for $u$ from \cite{Pat78} into
(\ref{eq:sigmf}). As a result,
\begin{equation}\label{eq:sigmagmf}
  \sigma = m^2.
\end{equation}
In zero magnetic field and for $K>K_c=1/q$ the model exhibits a
ferromagnetic transition. The spontaneous magnetization is given
by (\ref{eq:magmf}) with $h^{}_{\star}=0$: if $m^{}_{0}$ is a
solution, then $-m^{}_{0}$ is also a solution and both these
values minimize the free energy as function of the magnetization.
Once having determined the spontaneous magnetization, the relation
(\ref{eq:sigmagmf}) gives $\sigma^{}_{0}=m^{2}_{0}$. At constant
$K$, the ``van der Waals loops" of $h^{}_{\star}$ as function of
$m$ (for $|m|<|m^{}_{0}|$) are interpreted within mean-field-like
theories as non-equilibrium states: {\em metastable} for
$|m|>|m^{}_{s}|$ and {\em unstable} for $|m|<|m^{}_{s}|$, where
$\pm m^{}_{s}$ give the extremum points of $h^{}_{\star}$ (which,
at the same time, are the inflexion points of the free energy as a
function of the magnetization). By changing $K$, these two
extremum points describe the {\em spinodal} curve, and the
mean-field value of $m^{}_{s}$ is
\begin{equation}\label{magsmf}
  m^{}_{s}=\pm \sqrt{1-\frac{K_c}{K}} \ \ \mathrm{with} \ \
  K_c=\frac{1}{q}.
\end{equation}
The spin-spin correlation function along the spinodal curve is
then given by $\sigma^{}_{s}=m^{2}_{s}$.

\begin{figure}[p]
\begin{center}
\includegraphics[]{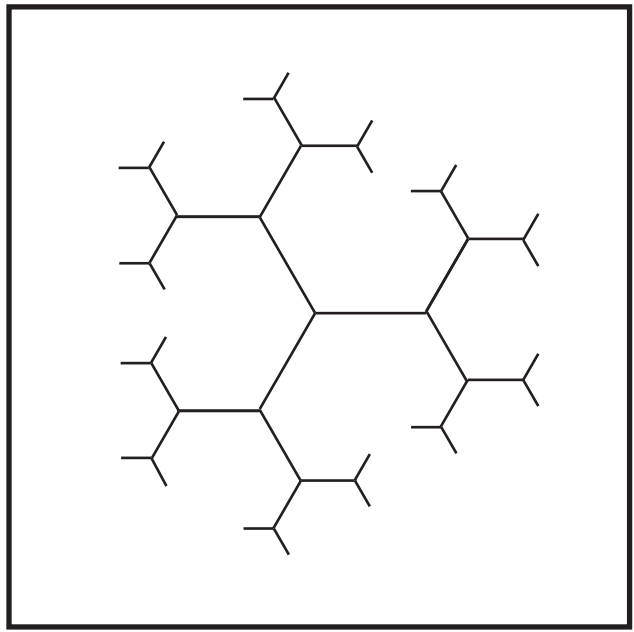}
\vspace{1cm} \caption{The Cayley tree with the coordination number
$q=3$ and the number of shells $n=4$. The Bethe lattice
corresponds to the limit $n\rightarrow \infty$.} \label{fig:bethe}
\end{center}
\end{figure}

In an improved scheme, with the same results as the Bethe
approximation \cite{Bax82}, we consider the Ising model on the
{\em Bethe lattice} (see Fig. \ref{fig:bethe}). Defining
\begin{equation}\label{eq:zmumu1}
  z=e^{-2K}, \; \mu=e^{-6h^{}_{\star}}, \; \mathrm{and} \;
  \mu^{}_{1}=\mu x^{q-1}_{},
\end{equation}
the magnetization of the Ising model on the Bethe lattice with the
coordination number $q$ is \cite{Bax82}
\begin{equation}\label{eq:magbl}
  m=\frac{1-\mu^{2}_{1}}{1+2z\mu^{}_{1}+\mu^{2}_{1}},
\end{equation}
while the spin-spin correlation function is \cite{Jed}
\begin{equation}\label{eq:sigbl}
  \sigma=\frac{1-2z\mu^{}_{1}+\mu^{2}_{1}}{1+2z\mu^{}_{1}+\mu^{2}_{1}},
\end{equation}
where $\mu^{}_{1}$ obeys the equation
\begin{equation}\label{eq:mu1/mu}
  \frac{\mu^{}_{1}}{\mu}=\left(
  \frac{z+\mu^{}_{1}}{1+z\mu^{}_{1}}
  \right)^{q-1}_{}.
\end{equation}
From (\ref{eq:magbl}) and (\ref{eq:sigbl}) we can get $\sigma$ as
a function of $z$ and $m$,
\begin{equation}\label{eq:sigmagbl}
 \sigma=1-2z\frac{1-m^{2}_{}}{z+\sqrt{1-(1-z^{2}_{})m^{2}_{}}}.
\end{equation}
This is the analog of Eq. (\ref{eq:sigmagmf}) of the mean-field
approximation and, for $z\rightarrow 1$ ($K\rightarrow 0$, i.e.,
small couplings), they coincide. In zero magnetic field and for
$z<z^{}_{c}=1-2/q$ the model undergoes a ferromagnetic transition.
Both $m^{}_{0}$ and $\sigma^{}_{0}$ can be determined in principle
by solving at first (\ref{eq:mu1/mu}) with $\mu =1$ and then
substituting the solution $\mu^{}_{1}(z)$ into (\ref{eq:magbl})
and (\ref{eq:sigbl}). In the particular case of $q=3$ one gets
\cite{HSB89}
\begin{equation}\label{eq:mag0bl}
 m^{}_{0}=\pm
 \frac{1}{1-2z}\left(\frac{1-3z}{1+z}\right)^{1/2}
\end{equation}
and
\begin{equation}\label{eq:sig0bl}
 \sigma^{}_{0}=1+2z\left[ \frac{1}{1-z}-\frac{1}{3(1+z)}-
 \frac{2}{3(1-2z)}\right].
\end{equation}
In determining the spinodal curve, we should first derive $\mu$ as
a function of $m$ using (\ref{eq:magbl}) and (\ref{eq:mu1/mu}),
and then one readily shows that $h^{}_{\star}$ has extremum points
at
\begin{equation}\label{eq:magsbl}
  m^{}_{s}=\pm \sqrt{\frac{1-z^{2}_{}/z^{2}_{c}}{1-z^{2}_{}}} \ \
  \mathrm{with} \ \ z^{}_{c}=1-\frac{2}{q}.
\end{equation}
Combining (\ref{eq:magsbl}) with (\ref{eq:sigmagbl}) we easily
find $\sigma^{}_{s}$.

\section{Phase diagrams}
\setcounter{equation}{0}

\hspace{\parindent}The ferromagnetic transition of the Ising model
occurs in zero magnetic field and for a coupling constant $K$
greater than a critical one $K_c$. This transition has a
correspondent in the molecular model only if $K_{\max}$ from
(\ref{eq:krange}), calculated at $h^{}_{\star}=0$, is greater than
$K_c$. The condition $\ln[\cosh(2L_1)]=2K_c$ then determines the
\emph{upper consolute temperature} $t^{}_{u}$ of the model, i.e.,
the maximum temperature at which a phase separation transition can
occur in the solution,
\begin{equation}\label{tu}
  t^{}_{u}=\frac{4}{\ln \left[ 2e^{2K_c}-1 \right]}.
\end{equation}
In the mean-field approximation of the Ising model ($q=3$) we get
$t^{}_{u}\simeq 3.762$; for the Bethe lattice, one has
$t^{}_{u}=4/\ln5 \simeq 2.485$.

\begin{figure}[p]
\vspace{-15cm} \hspace{0.5cm}
\includegraphics[]{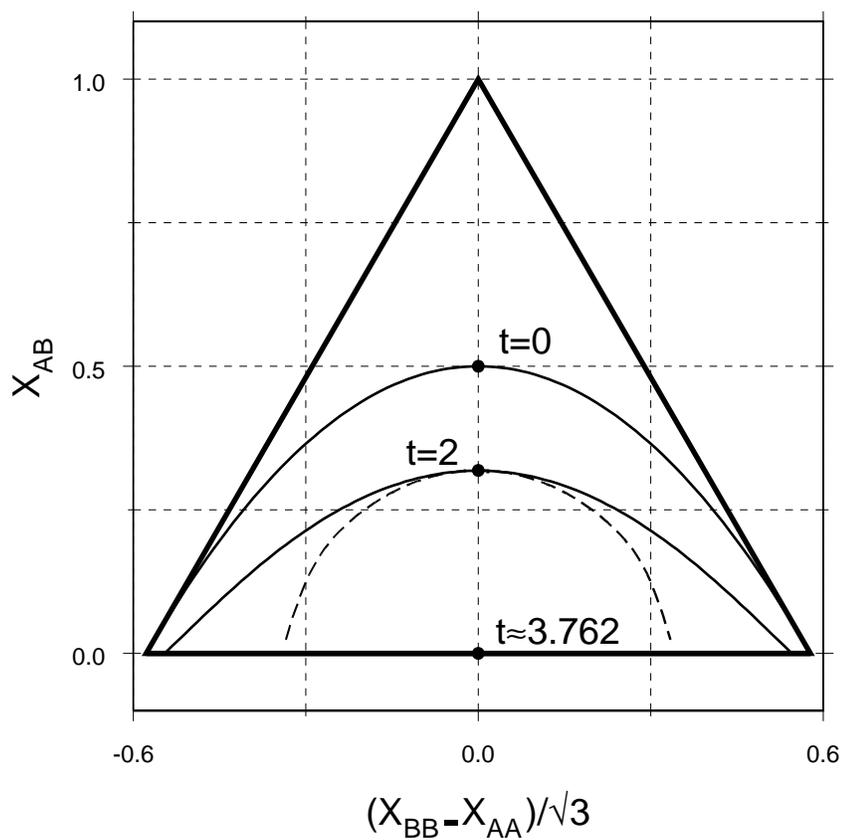}
\vspace{-2cm} \caption{Isothermal phase diagrams of the extended
Wheeler-Widom model on the honeycomb lattice when the equivalent
Ising model is treated in the mean-field approximation ($t$ is the
reduced temperature of the model). The coexistence curves are
drawn by solid lines, while the spinodal curves by dashed lines;
the solid circles mark the corresponding plait points. At $t=0$
the two curves (coexistence and spinodal) coincide.}
\label{fig:mf}
\end{figure}

In drawing the isothermal coexistence and spinodal curves, we use
the triangular diagram: a fixed composition is represented by a
point inside an equilateral triangle of the unit height, the mole
fraction of each component being given by the distance from that
point to the corresponding side of the triangle. For a triangular
diagram, the conservation equation (\ref{eq:consv}) is
automatically fulfilled, and the correspondence with the three
components of the molecular solution is chosen as shown in Fig.
\ref{fig:mf}.

Both the coexistence and the spinodal curves are symmetric under
reflections around the ($X_{AA}=X_{BB}$)-axis, a consequence of
the symmetry properties of the derivatives from (\ref{eq:sig312})
and (\ref{eq:mag312}) when $h^{}_{\star}\rightarrow -h^{}_{\star}$
(see \mbox{Appendix}). The analog of the Curie point for a
ferromagnet is the \emph{plait point}; it is given by the limit
$K\rightarrow K_c$ and it lies always on the
($X_{AA}=X_{BB}$)-axis. At the plait point, the spinodal curve is
tangent to the coexistence curve. In the vicinity of the plait
point, both the coexistence and the spinodal curves can be
approximated by parabolas. In the opposite limit, at $K=K_{\max}$,
it is shown in Appendix that $X_{AB}=0$ and, thus, both the
coexistence and spinodal curves reach the base of the triangular
diagram.

\begin{figure}[p]
\vspace{-15cm} \hspace{0.5cm}
\includegraphics[]{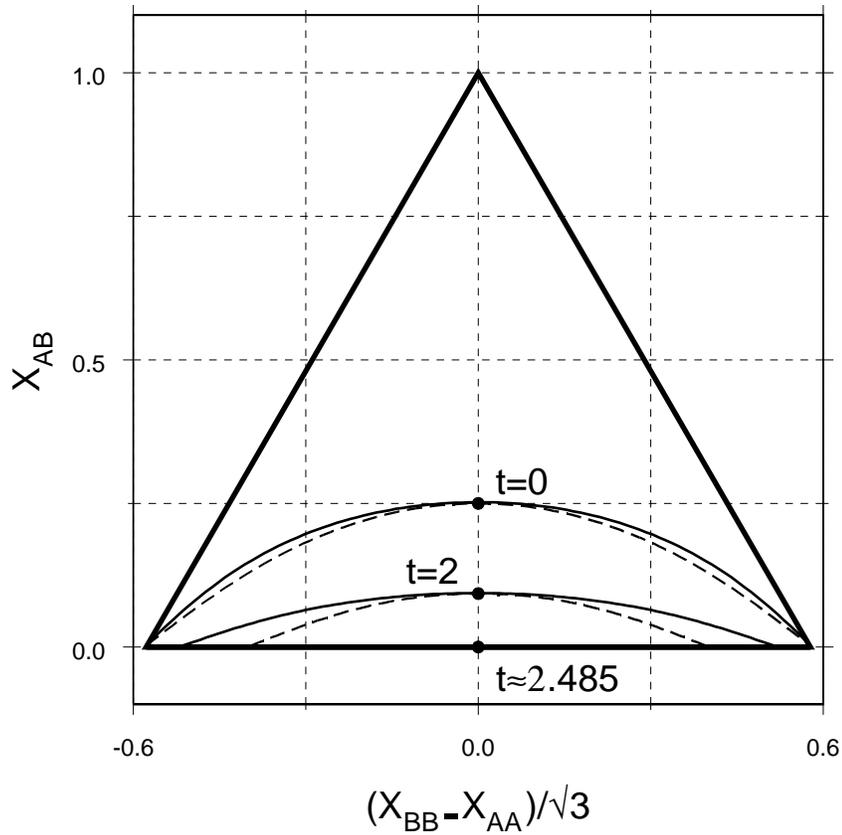}
\vspace{-2cm} \caption{The same as in Fig. \ref{fig:mf}, but in
the Bethe-lattice case. At $t=0$ the spinodal curve does
\emph{not} coincide with the coexistence curve.} \label{fig:bl}
\end{figure}

\begin{figure}[p]
\vspace{-15cm} \hspace{0.5cm}
\includegraphics[]{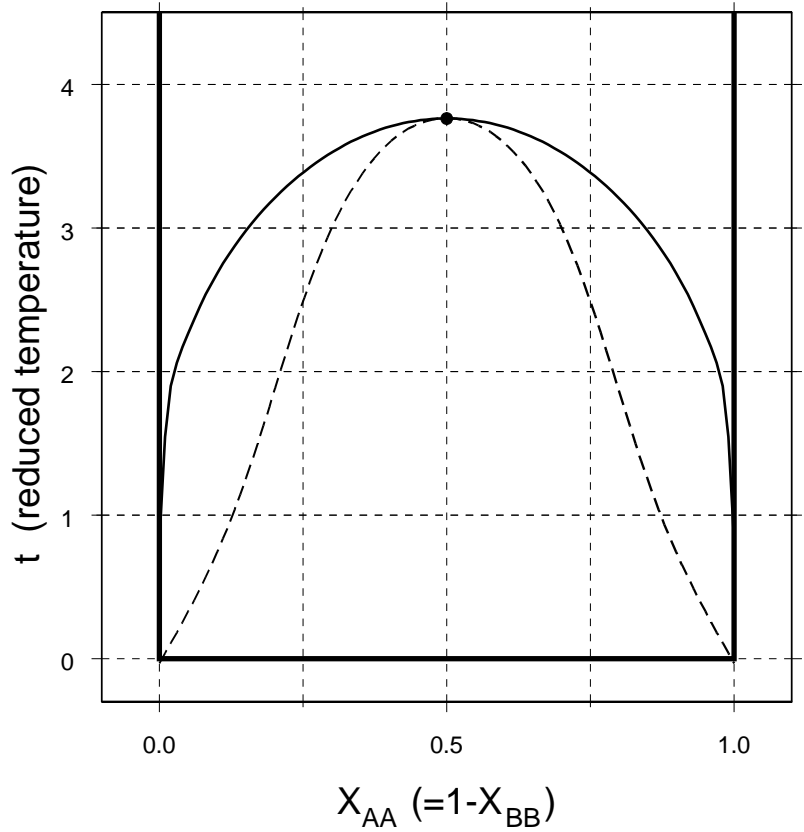}
\vspace{-2cm}
\caption{The phase diagram of the extended
Wheeler-Widom model on the honeycomb lattice in the limit of two
components ($X_{AB}=0$) and when the equivalent Ising model is
treated in the mean-field approximation. The coexistence curve is
represented by a solid line, the spinodal by a dashed line, and
the plait point by a solid circle.} \label{fig:mf2}
\end{figure}

\begin{figure}[p]
\vspace{-15cm} \hspace{0.5cm}
\includegraphics[]{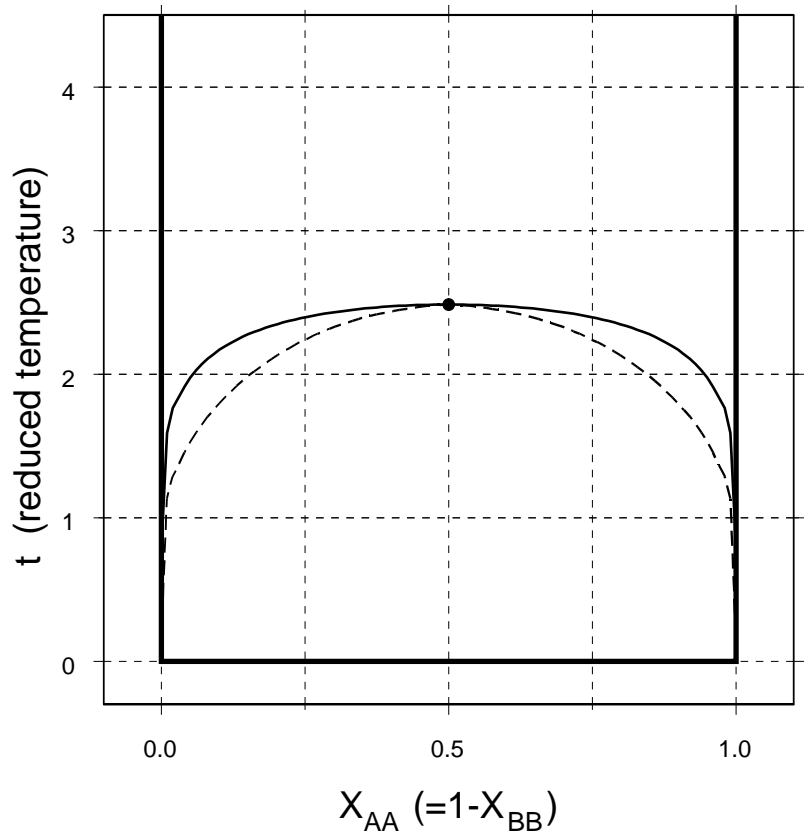}
\vspace{-2cm} \caption{The same as in Fig. \ref{fig:mf2}, but in
the Bethe-lattice case.} \label{fig:bl2}
\end{figure}

The phase diagrams at some particular values of the reduced
temperatures are presented in Figs. \ref{fig:mf} and \ref{fig:bl},
respectively corresponding to the mean-field and Bethe-lattice
approximations of the equivalent Ising model. In the mean-field
approximation (Fig. \ref{fig:mf}), at $t=0$ the spinodal and the
coexistence curve coincide; they are described by the same
parabola $y=(1-3x^2)/2$. This means that no metastable states are
predicted at zero temperature, an expected result from more
general considerations. Technically, this is a consequence of the
relation (\ref{eq:sigmagmf}) between $\sigma$ and $m$, which is
independent of the coupling constant $K$: the distinct
$K$-dependencies of $m$ for the spinodal and coexistence curves
play only a scaling role (the same curve is covered in different
ways). In the Bethe-lattice case (Fig. \ref{fig:bl}), the relation
(\ref{eq:sigmagbl}) between $\sigma$ and $m$ explicitly depends on
the coupling constant, and, in general, it should be so.
Consequently, the three-component system can be in a metastable
state even at zero temperature, a fact that might reveal an
inconsistency of the theory. (The implications of this result and
a possible way to overcome the difficulty within mean-field-like
theories of first-order phase transitions will be discussed in a
forthcoming paper.) By increasing the temperature, the curves go
down and shrink to a point when $t=t^{}_{u}$. The limiting case
$X_{AB}=0$ corresponds to a two-component system. In the
mean-field approximation of the equivalent Ising model, the
spinodal and the coexistence curves depend on temperature as in
Fig. \ref{fig:mf2}; the case of the Bethe lattice is illustrated
in Fig. \ref{fig:bl2}.

\section{Discussion}

\hspace{\parindent}The Wheeler-Widom model (with finite, two-body
interactions) for three-component molecular systems is equivalent
to the standard (spin-1/2, nearest-neighbor interaction) Ising
model. Based on this equivalence, we derived the spinodal curve
for the molecular model on a honeycomb lattice using approximate
(mean-field and Bethe-lattice) results for the Ising model. Both
the spinodal and the (previously determined) coexistence curves
have been drawn at different temperatures.

The present work can be extended in several ways. One direction is
to consider \emph{three-body} interactions between the molecular
ends associated with the same site of the honeycomb lattice. The
coexistence curves then become \emph{asymmetric}, with the plait
point located off the ($X_{AA}=X_{BB}$)-axis of the triangular
diagram \cite{SHW91}; a similar asymmetry should be observed for
the corresponding spinodal curves. Another possible development is
to determine the spinodals for systems with both upper and
\emph{lower} consolute temperatures. Within the present formalism,
a lower bound on the transition temperature follows from the
assumption that the molecular ends may form \emph{bonds}; the
coexistence surface for a three-coordinate Bethe lattice has been
determined in \cite{HuS89}. A more general situation, with two
possible transitions (one with both upper and lower consolute
temperatures, the second one only with an upper consolute
temperature below the minimum transition temperature of the first
one) has been analyzed in \cite{HuS90}. It is also possible to
generate phase diagrams with all the characteristics mentioned
above (asymmetric, upper and lower consolute temperatures, etc.)
within a single ternary solution model of the Wheeler-Widom type
\cite{BuH01}.

In the framework of the formalism presented above, the spinodal
curve of a ternary solution is determined starting from a simple
microscopic model reducible to two components. Together with
analysis of experimental data, it could contribute to a better
understanding of interesting phenomena such as diffusion and
nucleation in multi-component systems. Another important aspect
should also be mentioned: since a Wheeler-Widom-type model is
equivalent to the standard Ising model, it perhaps provides the
only way to get exact results for systems with three components.

\subsection*{Acknowledgements} \hspace{\parindent}
We thank Dale A. Huckaby for many valuable discussions on this
subject. The support of the NASA Microgravity Biotechnology
Program through Grant NAG8-1356 is gratefully acknowledged.

\section*{Appendix}

\setcounter{equation}{0}
\renewcommand{\theequation}{A.\arabic{equation}}
\hspace{\parindent} Let us denote the derivatives that appear in
(\ref{eq:sig312}) and (\ref{eq:mag312}) by
\begin{equation}\label{eq:cijdef}
  \left\{
  \begin{array}{ll}
   {\displaystyle c_{11}= \left(\frac{\partial \ln B}{\partial L}\right)_{L_{1},h}} & , \ \
   {\displaystyle c_{21}=\frac{1}{2}\left(\frac{\partial \ln B}{\partial h}\right)_{L_{1},L}} \\
     & \\
   {\displaystyle c_{12}=\left(\frac{\partial K}{\partial L}\right)_{L_{1},h}} & , \ \
   {\displaystyle c_{22}=\frac{1}{2}\left(\frac{\partial K}{\partial h}\right)_{L_{1},L}} \\
     & \\
   {\displaystyle c_{13}=2\left(\frac{\partial h_{\star}}{\partial L}\right)_{L_{1},h}} & , \ \
   {\displaystyle c_{23}=\left(\frac{\partial h_{\star}}{\partial h}\right)_{L_{1},L}}.
  \end{array} \right.
\end{equation}
Using the notations
\begin{equation}\label{eq:pz}
 p=\left( e^{4R}+1 \right) /2 \ \ \mathrm{and} \ \ z=e^{-2K},
\end{equation}
the quantities $\chi$ and $\lambda$ respectively defined in
(\ref{eq:chi}) and (\ref{eq:lambda}) become
\begin{equation}
 \chi=\frac{\sqrt{p^2-1}\sinh(2h^{}_{\star})}{p\cosh(2h^{}_{\star})-z}
 \ \ \mathrm{and} \ \
 \lambda=\frac{p\cosh(2h^{}_{\star})-z}{pz-\cosh(2h^{}_{\star})}.
\end{equation}
Introducing
\begin{equation}
 \omega=(1+p\lambda)^2-(p^2-1)\chi^2\lambda^2,
\end{equation}
the $c$-quantities from (\ref{eq:cijdef}) are given in terms of
$K$, $h^{}_{\star}$, and $R$ as
\begin{equation}\label{eq:cij}
  \left\{
  \begin{array}{ll}
   {\displaystyle c_{11}=\frac{\lambda}{p+\lambda}-\frac{1+p\lambda}{\omega}} & ,
   \ \ {\displaystyle c_{21}=\frac{1+z^2}{2}\frac{\chi \lambda}{p+\lambda}}
   \\ & \\
   {\displaystyle c_{12}=\frac{p}{p+\lambda}-\frac{1+p\lambda}{\omega}} & ,
   \ \ {\displaystyle c_{22}=-\frac{1-z^2}{2}\frac{\chi \lambda}{p+\lambda}}
   \\ & \\
   c_{13}=2\sqrt{p^2-1} \ {\displaystyle \frac{\chi\lambda}{\omega}} & ,
   \ \ c_{23}=\sqrt{p^2-1} \ [1+p(1-\chi^2)\lambda]{\displaystyle
   \frac{\lambda}{\omega}}.
  \end{array} \right.
\end{equation}

Taking the limit $R\rightarrow \infty$ in (\ref{eq:cij}), we get
$c_{11}=c_{13}=c_{21}=c_{22}=0$ and $c_{12}=c_{23}=1$. This limit,
understood as $\varepsilon_{AB}\rightarrow \infty$ [see
(\ref{eq:r})], corresponds to the original model introduced by
Wheeler and Widom \cite{WhW68}.

Let us remark that $\chi$ is an odd function of $h^{}_{\star}$,
while $\lambda$ and $\omega$ are even functions of $h^{}_{\star}$
[see (A.3) and (A.4)]. As a consequence, the quantities $c_{11}$,
$c_{12}$, and $c_{23}$ are symmetric under the transformation
$h^{}_{\star}\rightarrow -h^{}_{\star}$, whereas $c_{13}$,
$c_{21}$, and $c_{22}$ are antisymmetric. Because $m$ is
antisymmetric in $h^{}_{\star}$ and $\sigma$ is an even function
of $m$ (see \mbox{Section 3}), from the mole fractions equations
(\ref{eq:consv}) to (\ref{eq:molfrmag}) it follows that $X_{AB}$
is symmetric in $h^{}_{\star}$, while $X_{AA}-X_{BB}$ is
antisymmetric. These properties determine the symmetry of the
phase diagrams (discussed in Section 4) under reflections around
the ($X_{AA}=X_{BB}$)-axis. Moreover, at $h^{}_{\star}=0$ all
$c_{13}=c_{21}=c_{22}=0$ and
$\sigma^{}_{\mbox{\scriptsize{3-12}}}$ decouples from
$m^{}_{\mbox{\scriptsize{3-12}}}$ [see (\ref{eq:sig312}) and
(\ref{eq:mag312})].

Another interesting limit is when $K\rightarrow K_{\max}$ [see
(\ref{eq:krange})]. In this case $\lambda \rightarrow \infty$,
$\omega$ goes to infinity as $\lambda^2$, and from (\ref{eq:cij})
we get $c_{11}=1$, $c_{12}=c_{13}=0$; consequently,
$\sigma^{}_{\mbox{\scriptsize{3-12}}}$ given by (\ref{eq:sig312})
is equal to one. From (\ref{eq:consv}) and (\ref{eq:molfrsig}) it
then follows that $X_{AB}=0$. This property is used in Section 4
to show that in the limit $K\rightarrow K_{\max}$ the coexistence
and the spinodal curves always reach the bottom side of the
triangular diagram.

\end{document}